\documentclass[16pt,a4]{article}
\usepackage{epsfig}
\usepackage{setspace}
\title{Deposition of Diamond-like Carbon films using Dense Plasma Focus}
\author{Chhaya Ravi Kant$^a$, P.Arun$^b$, Savita Roy$^c$ \&
M.P.Srivastava$^d$\\
$^a$Department of Applied Sciences,\\
Indira Gandhi Institute of Technology\\
Guru Gobind Singh Indraprastha University,\\ Delhi 110 006, India.\\
\\
$^b$Department of Physics and Electronics,\\
S.G.T.B. Khalsa College\\
University of Delhi, Delhi 110 007, India.\\
\\
$^c$Department of Physics\\
Kalindi College\\
University of Delhi, Delhi 110 007, India.\\
\\
$^d$Department of Physics \& Astrophysics\\
University of Delhi, Delhi 110 007, India.\\
}

\begin{document}
\maketitle
\begin{abstract}
Diamond Like Carbon (DLC) films were deposited on quartz substrates using 
Dense Plasma Focus (DPF) method. The formation of ${\rm sp^3}$ bonds as is 
it's content in the film strongly 
depends on the substrate and it seems quartz is not a suitable substrate. 
However, we report here the formation of DLC films on quartz substrates that
were maintained at elevated temperatures and show the content of carbon atoms 
with ${\rm sp^3}$ bonds in the film is directly proportional to the
substrate's temperature. Not only does this give good control of film 
fabrication, but also shows how to take advantage of DLC film's anti-wear, 
scratch resistant properties on material surfaces that require protection. 
\end{abstract}
{\bf PACs No: 81.05.Uw, 78.55.Qr, 78.20.-e}
\vfil \eject

\section*{Introduction}

Carbon exists in various allotropic forms in nature of which the two most 
popular states are the trigonally bonded (${sp^2}$) graphite form and the 
tetragonal bonded (${sp^3}$) diamond form. Since the last decade, the 
amorphous mixture containing significant 
fraction of diamond like bonded carbon intimately mixed with graphite has been
studied. Due to the unique features exhibited by this mixture, this form is 
now catagorized as {\sl Diamond like Carbon (DLC)}. The term DLC was first 
used by Aisenberg and Chabot \cite{r2}. In contrast to amorphous carbons, 
this new form of carbon is hard, electrically insulating and wear resistant. 
That is, though DLC exhibits Diamond like properties it lacks the crystallinity
of diamond.

Due to its unique structural, mechanical, optical and electronic properties, 
DLC films have wide range of potential applications \cite{r2a}. They are used 
for anti-reflection and protective coatings in IR and UV optics and 
tribological applications such as a mechanically hard, scratch resistant 
layer to coat cutting tools so as to protect them against corrosion and wear. 
DLC films also have potential application for electronic packaging, 
passivation and thermal heat sinks for high power devices because of their 
high thermal conductivity and electrical resistivity. Being biocompatible, 
wear resistant, and chemically inert, DLC films are also used to coat heart 
valves, rods inserted inside human limbs and contact lenses. Interestingly,
the properties of this form of carbon is strongly dependent on the 
${sp^3/sp^2}$ ratio in the mixed structure.  

The potential applications of DLC films have stimulated a great deal of 
interest in present decade. Various methods have been developed to 
deposit DLC films, like DC sputtering \cite{r4}, RF sputtering \cite{r5}, 
magnetron sputtering \cite{5a,6}, ion beam sputtering \cite{7,8}, plasma 
assisted physical vapour deposition \cite{9}, plasma assisted chemical vapour 
deposition \cite{10}, ion beam deposition \cite{r2},\cite{11}-\cite{13}, fast 
atomic beam bombardment (FAB) technique \cite{14}, carbon arc evaporation
\cite{15,16}, laser evaporation \cite{17,18}, (VHF PECVD) technique
\cite{19} and more recently by Dense Plasma Focussing (DPF) \cite{dpf1, dpf2}.

A common feature of all the methods mentioned above is that the
DLC film is condensed from a beam containing medium energy $\sim$100eV
carbon or hydrocarbon ions.  
In this paper we discuss the deposition of DLC films by a 
Dense Plasma Focus (DPF) device where the beam energy is considerably
higher. In the following passages we report the
results of various characterisation done on the DLC films deposited by DPF. 
Also, it is worth
mentioning here that in the present study we have used quartz substrates 
unlike the earlier reports on DLC film fabrication
using DPF \cite{dpf1, dpf2} where silicon wafers were
used as substrates.

\section{Experimental Details}

We have used a Mather type DPF device of available in the Plasma research 
laboratory of Delhi University. The geometry of the DPF 
Device and its
working is detailed elsewhere \cite{20}-\cite{26}. The Dense Plasma Focus Device (DPF) produces 
energetic (1-2 keV) high-density plasma for duration of 
about hundred nanoseconds \cite{17, 18, 20}. The chamber was evacuated and 
filled with Argon gas with the pressure inside 
the chamber maintained in the range of 80-120 Pa. Energetic ions
move from source to substrate in a cylindrical geometry along the axial 
direction of central electrode. These highly energetic carbon ions were 
deposited on quartz substrates placed at the top of the chamber. DLC films
were thus obtained on quartz substrates maintained at 
different temperatures.

The deposition rate of our apparatus in the present study (arrangement) was
found to be 450\AA\, per shot. All our samples were grown with four focused 
shots of carbon ions from Dense Plasma Focus device. 
Each DPF shot was of 100ns duration. Thus, the average thickness of the our 
films were 1800\AA. The thickness of the films were measured and confirmed 
using a Dektak profilometer. The profilometer also showed the films 
to have a smooth topography with low surface roughness. 

The structural and morphological analyses of our films were carried out. 
X-Ray diffraction studies of the films were carried out using Phillips 
PW-1840 X-ray diffractometer. The SEM micrographs of our 
films, taken using JEOL JSM-1840 Scanning Electron Microscope.

\section{Results and Discussions}

Visual examination showed the film grown at room temperature to be blackish 
in color. This is characteristic of graphite films. However, films grown on 
substrates maintained between temperatures ${\rm 100^oC}$ and ${\rm 300^oC}$ 
were found to be transparent. The transparent films were found to be 
scratch resistant and chemically inert. Chemical inertance is based on our 
test of applying drops of 
concentrated hydro-chlorine acid (HCl) and nitric acid (${\rm HNO_3}$) on the
film's surface. The films showed remarkable stability. Also, the films
showed no sign of disintegration or peeling even 
after months of deposition. This stability and the transparent nature of the 
films grown on
quartz substrates kept at elevated temperatures indicate the formation of
DLC films. In the following sections we report the structural and optical
characterisation of these samples which would prove the formation of DLC
films.

\subsection{Structural and Morphological Studies}

As expected the diffractograms of the films grown at room temperature and at 
higher temperature showed them to be amorphous in nature without exception 
(fig~1). The hump seen in the X-ray diffractogram however is indicative of
some short range ordering among the atoms of the films. The morphological
study is indicative of this with no features evident on the films surface
when examined under scanning electron microscope (fig~2a). The blackish
asgrown films showed granularity with grain sizes of the order of 
$\sim$100nm. The Atomic force microscope (AFM) image of the samples surface
confirms the surfaces granularity (fig~3a). One would expect X-Ray 
diffraction peaks if 
ordering of atoms takes place over larger distances accompaning increase in 
grain size. However, the surface of the transparent films obtained by growing 
them on heated substrates did not exhibit features similar to that obtained
in asgrown films (fig~2b). The grains are small, sparse and scattered as can
also be seen from the AFM images of the films grown on heated substrates
(fig~3). Also,
notice the surface roughness is minimum for the film grown on substrates
maintained at ${\rm 150^oC}$. We believe the surface roughness may be an
important contributing factor in the optical properties of the films that we
investigate in the following section.

\section{Optical Properties}
\subsection{Raman Studies}

The Raman spectra have been recorded at room temperature with 514.5nm line 
of a spectra Physics 2030 ${\rm Ar^+}$ laser in the backscattering geometry.
A laser beam of 5mW power was used. The spectra were recorded with a Spex 
Triplemate recorder. Fig~4 shows the Raman spectra of the DLC films deposited 
by DPF device on quartz substrates kept at ${\rm 100^oC}$, ${\rm 150^oC}$ and 
${\rm 300^oC}$
respectively. The broad peak ${\rm 1580cm^{-1}}$ and ${\rm 1350cm^{-1}}$ 
corresponding to the G band and D band overlap and have been deconvoluted. 
Deconvolution enables us to estimate the Peak Position, Area, Maximum
Intensity and the Full Width at Half Maxima (FWHM) of each band. The results
are listed in Table~1. The G peak's FWHM is a measure of disorder of bond
angles in graphites. This is a measure of the stress in the samples.
Robertson \cite{JAP} states that the Raman peak's position decreases linearly 
with increasing stress. In other words,
\begin{eqnarray}
Peak \quad Position \propto {1 \over FWHM}\nonumber
\end{eqnarray} 

%%%%%%%%%%%%%%%%%%%%%%%%%%%%%%%%%%%%%%%%%%%%%%%%%%%%%%%%%%%%%%%%%%%%%%%%%%
\begin{table}[h]
\begin{center}
\caption{\sl Table compares the D and G peaks of the Raman Spectra for films
grown on quartz substrates maintained at ${\rm 100^oC}$, ${\rm 150^oC}$ and 
${\rm 300^oC}$
respectively.}
\vskip 0.5cm
\begin{tabular}{|c| c|c|c|c| c|c|c|c|}
\hline
 &\multicolumn{4}{c|}{D line}&\multicolumn{4}{c|}{G line} 
\\ 
%\hline
Temp.&Center & Int.& Area & FWHM & Center &Int.& Area & FWHM\\ \hline
${\rm 100^oC}$ & 1340 & 1509 & ${\rm 6e^5}$ &256.6 & 1600 & 1912 & ${\rm
2e^5}$ & 68.7 \\ \hline
${\rm 150^oC}$ & 1339.6 & 3063 & ${\rm 1e^6}$ & 221 & 1594.6 & 3925 & ${\rm
4e^5}$ & 73.6 \\ \hline
${\rm 300^oC}$ & 1357.7 & 8567 & ${\rm 3.37e^6}$ & 251.06 & 1584.2 & 9837 & 
${\rm 1.56e^6}$ & 100.94 \\ \hline
\end{tabular}
\label{tab1}
\end{center}
\end{table}
%%%%%%%%%%%%%%%%%%%%%%%%%%%%%%%%%%%%%%%%%%%%%%%%%%%%%%%%%%%%%%%%%%%%%%%%%%

This is true in our samples, where there is an increase in the G peak's FWHM
accompanied with a decrease in the peak position as the substrate's
temperature is raised. This implies that with increased substrate
temperature, the stress in the film (in terms of disorder in graphite bond
angles) increased. Beeman modelled \cite{bee} DLC films and predicted 
that the  
increasing diamond like structure generates stress within the film. That is,
the shifting of the `G' peak is also an indication of the formation of ${\rm
sp^3}$ bonds. 

The ${\rm sp^3}$ content have also be estimated from the ratio of 
areas enclosed by the two peaks. The ratio 
${\rm I(D)/I(G)}$ is also said to be inversely proportional to the fractional 
ratio of
${\rm sp^3}$ bonded material present to that of ${\rm sp^2}$ bonded carbon
\cite{JAP}. However, in our
case ${\rm I(D)/I(G)}$ decreased from ${\rm \approx 0.79}$ to
0.78 for increase in substrate temperature from ${\rm 100^oC}$ to ${\rm
150^oC}$ and subsequently increased to 0.87 for films grown on substrates
maintained at ${\rm 300^oC}$. This suggests an increase in ${\rm sp^3}$
content followed by a decrease with substrate heating. Singha \cite{singha}
et al however consider the ratio ${\rm I(D)/I(G)}$ to be a poor measure of
the ${\rm sp^3}$ content.

Ferrari and Robertson \cite{ferrari} have characterised the formation of 
DLC in three stages (a) stage 1 that of NC carbon, (b) stage 2 that of
a-carbon and (c) stage 3 of ta-carbon. Each stage is well identifiable using
their Raman spectras. As per this model our DLC samples are of stage 2, 
having their G peak position between 1600 to 1510${\rm cm^{-1}}$ and the 
ratio of peak intensities, ${\rm I(D)/I(G)}$ between 0.25 to 2. Stage 2
films are also marked by `G' peaks with FWHM greater than ${\rm 50cm^{-1}}$
indicative of grain size of the order of 1nm. AFM images (fig~3) also
substantiate that our samples can be classified in stage 2. Using data
from Ferrari and Robertson \cite{ferrari} and that of Tamor et al
\cite{tamor}, Singha et al \cite{singha} gives the content of ${\rm sp^3}$
in DLC samples of stage 2 as
\begin{eqnarray}
Content (sp^3)=0.24-48.9(\omega_G-1580)\times 10^{-4}\label{e1}
\end{eqnarray}
where ${\rm \omega_G}$ is the position of the Raman `G' peak. Calculating
the percentage of ${\rm sp^3}$ in our samples using eqn(\ref{e1}) and data
listed in Table I, we find that content of ${\rm sp^3}$ increased from
14.22\% to 22\% as the substrate temperature was increased. A plot of ${\rm
sp^3}$ content with respect to substrate temperature (fig~5) shows perfect
linearity. Thus, good control over the quality of DLC films can be obtained
by controlling the substrate temperature.

To compare contribution of substrate on DLC formation, we placed a silicon
substrate along with the quartz substrate and maintained both at ${\rm
300^oC}$. Deposition of films were done simultaneously and yet the film
grown on silicon (fig~6) had 30.36\% of carbon atoms with ${\rm sp^3}$
bondings as compared to just 22\% obtained on the quartz substrates
(fig~4c). This maybe a result of dangling bonds of silicon on the wafer
surface acting as center of seeding, encouraging formation of ${\rm sp^3}$
bondings. 

\subsection{UV Spectroscopy}

The transmission spectra of the DLC films were obtained using Shamadzu UV-260 
spectrophotometer. Fig~7 shows the variation of transmittance (percentage) 
of the films deposited at various substrate temperatures. As expected the
graphite films grown at room temperature with it's opaqacity has the lowest 
transmittance. The nature of the curves in fig~7 suggests that the band gaps
of the flms lie in the UV
region. Hence, we have not attempted to calculate Tauc's bandgap from this
data. 

The films grown at ${\rm 100^oC}$ and ${\rm 150^oC}$ have
very similar amount of atoms in
${\rm sp^3}$ bondings and yet their transmittance is sharply different. We
believe the surface smoothness of samples grown at ${\rm 150^oC}$ show
better transmission due to
lower surface scattering. As the
surface roughness increases (for sample `d', grown on substrate kept at
${\rm 300^oC}$), the transmittance decreases. 

\section{Conclusion}
To conclude, we have deposited Diamond like Carbon films using Dense Plasma 
Focus Device on quartz substrate at moderate temperatures. Till now DLC
films have been fabricated by various methods on Silicon substrates by
varying the beam energy. This technique shows that good quality films can be
grown on different substrates and simple method of varying substrate
temperature not only gives DLC films but also gives a control over the
optical properties of the resulting film. This gives true meaning to the 
tribological applications of DLC films where it can be grown on any surface
that requires mechanical/ scratch resistant properties.

\section{Acknowledgment}

We are thankful to Mr. N.C. Mehra and Raman 
Malhotra University
Science and Instrumentation Center (USIC), Delhi University, for carrying
out the analysis of our samples.

\vfil
\section*{Figure Captions}
\begin{itemize}

\item[1] X-ray diffractogram of DLC film prepared on substrates maintained at 
${\rm 300^oC}$.

\item[2] Scanning Electron Micrograph of DLC film prepared at (a) room
temperature and (b) on substrates maintained at ${\rm 300^oC}$.

\item[3] Atomic Force Microscope images of DLC film prepared at (a) room
temperature, (c) on substrates maintained at ${\rm 100^oC}$, (e) on 
substrates maintained at ${\rm 150^oC}$ and (g) on substrates maintained at
${\rm 300^oC}$. Images (b), (d), (f) and (h) are the 3-D view giving an idea
of the film surface's roughness shown in (a), (c), (e) and (g) respectively.

\item[4] Raman Spectra of DLC films deposited on quartz substrates
maintained at temperature (a) ${\rm 100^oC}$, (b) ${\rm 150^oC}$ and (c) 
${\rm 300^oC}$. The 
deconvulated curves represents the D and G peaks.

\item[5] The content of carbon atoms with ${\rm sp^3}$ bondings increases
linearly with temperature at which the quartz substrates were maintained.

\item[6] Raman Spectra of DLC films deposited on Silicon substrates
maintained at ${\rm 300^oC}$. The 
deconvulated curves represents the D and G peaks. The peak between
900-1000${\rm cm^{-1}}$ is of silicon from the substrate.
    
\item[7] The UV-visisble transmission spectra of DLC films deposited on 
quartz substrates maintained at temperature (a) room temperature, (b) 
${\rm 100^oC}$, (c) ${\rm 150^oC}$ and (d) ${\rm 300^oC}$.

\end{itemize}

\vfil \eject
\section*{Figures}

%%%%%%%%%%%%%%%%%%%%%%%%%%%%%%%%%%%%%%%%%%%%%%%%%%%%%%%%%%%%%%%%%%%%%%%%%%%
\begin{figure}[h!!]
\begin{center}
\includegraphics[width=2.25in,angle=-90]{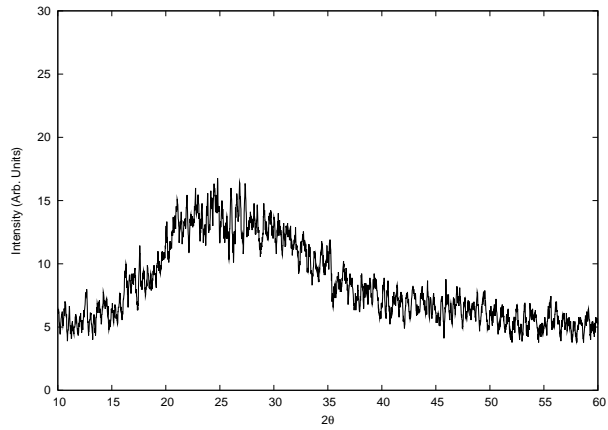}
\caption{X-ray diffractogram of DLC film prepared on substrates maintained at 
${\rm 300^oC}$.}
\end{center}
\end{figure}
%%%%%%%%%%%%%%%%%%%%%%%%%%%%%%%%%%%%%%%%%%%%%%%%%%%%%%%%%%%%%%%%%%%%%%%%%%

%%%%%%%%%%%%%%%%%%%%%%%%%%%%%%%%%%%%%%%%%%%%%%%%%%%%%%%%%%%%%%%%%%%%%%%%%%%
\begin{figure}[h!!]
\begin{center}
\includegraphics[width=2.25in,angle=0]{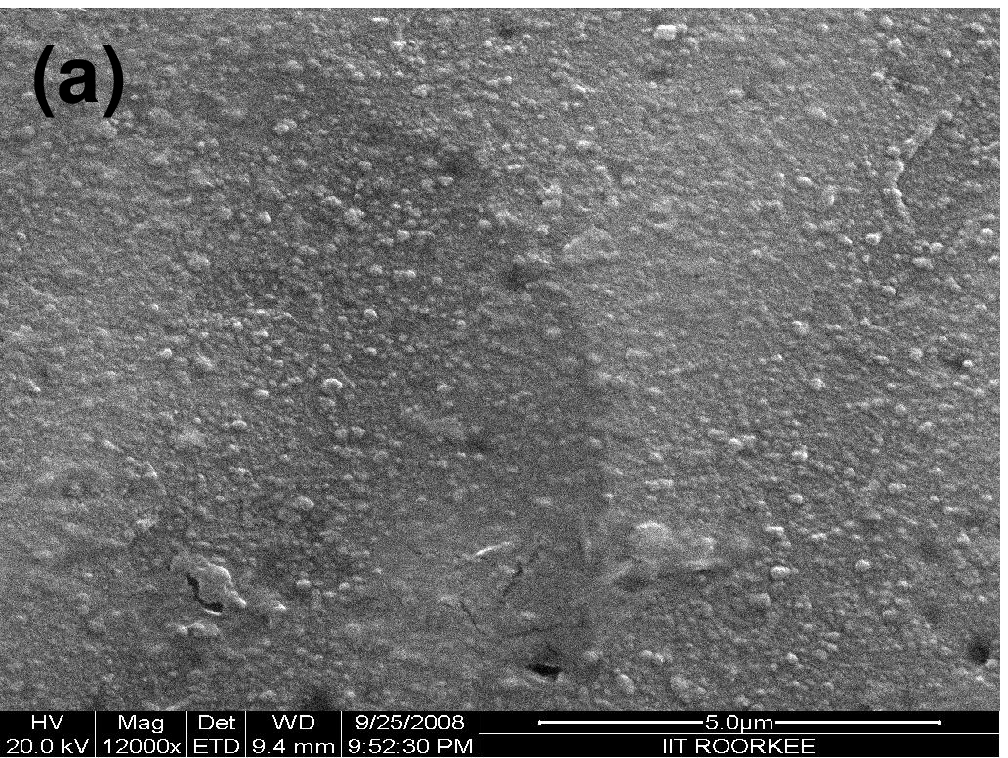}
\hfil
\includegraphics[width=2.25in,angle=0]{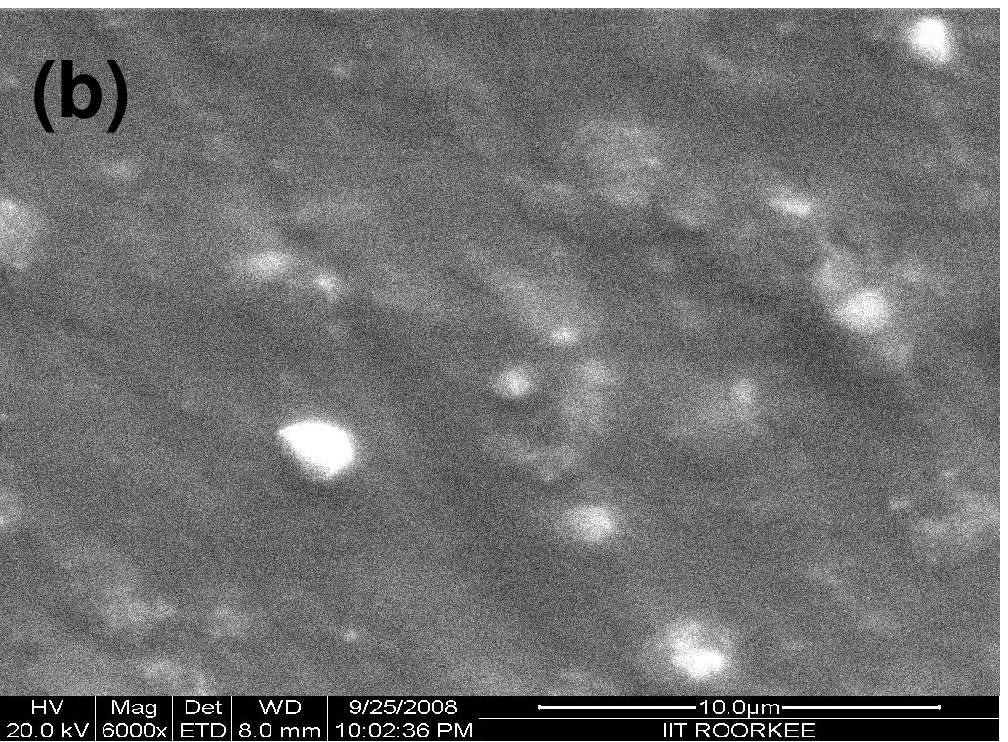}
\caption{Scanning Electron Micrograph of DLC film prepared at (a) room
temperature and (b) on substrates maintained at ${\rm 300^oC}$.}
\end{center}
\end{figure}
%%%%%%%%%%%%%%%%%%%%%%%%%%%%%%%%%%%%%%%%%%%%%%%%%%%%%%%%%%%%%%%%%%%%%%%%%%

%%%%%%%%%%%%%%%%%%%%%%%%%%%%%%%%%%%%%%%%%%%%%%%%%%%%%%%%%%%%%%%%%%%%%%%%%%%
\begin{figure}[t!!]
\begin{center}
\includegraphics[width=2.7in,angle=0]{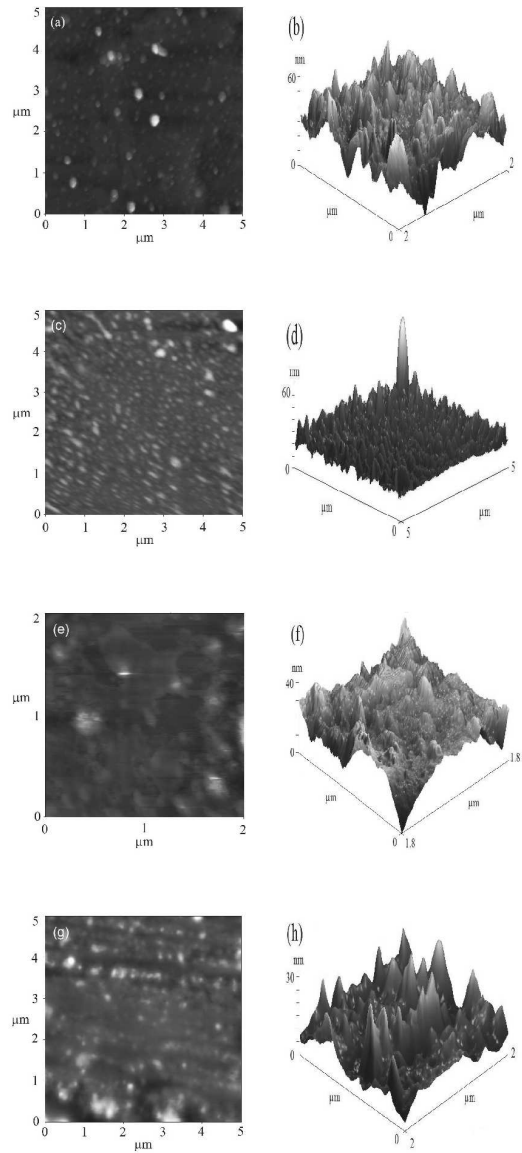}
\caption{Atomic Force Microscope images of DLC film prepared at (a) room
temperature, (c) on substrates maintained at ${\rm 100^oC}$, (e) on 
substrates maintained at ${\rm 150^oC}$ and (g) on substrates maintained at
${\rm 300^oC}$. Images (b), (d), (f) and (h) are the 3-D view giving an idea
of the film surface's roughness shown in (a), (c), (e) and (g) respectively.}
\end{center}
\end{figure}
%%%%%%%%%%%%%%%%%%%%%%%%%%%%%%%%%%%%%%%%%%%%%%%%%%%%%%%%%%%%%%%%%%%%%%%%%%

%%%%%%%%%%%%%%%%%%%%%%%%%%%%%%%%%%%%%%%%%%%%%%%%%%%%%%%%%%%%%%%%%%%%%%%%%%%
\begin{figure}[t!!]
\begin{center}
\includegraphics[width=1.7in,angle=-90]{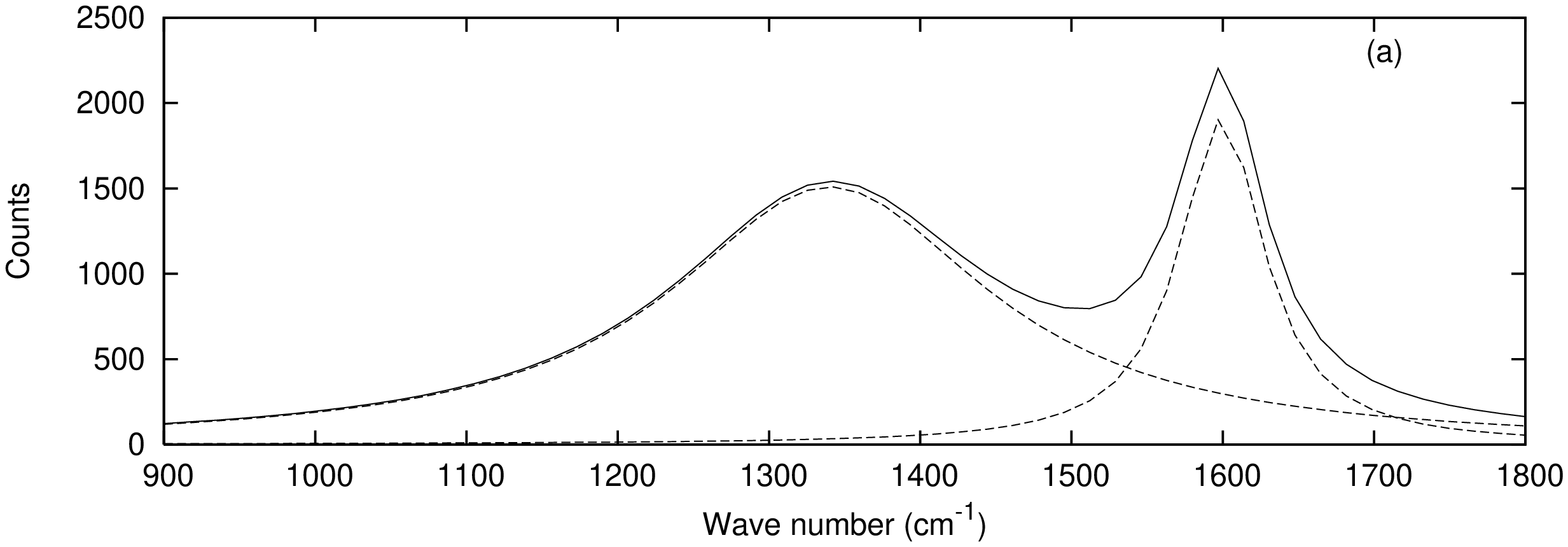}
\includegraphics[width=1.7in,angle=-90]{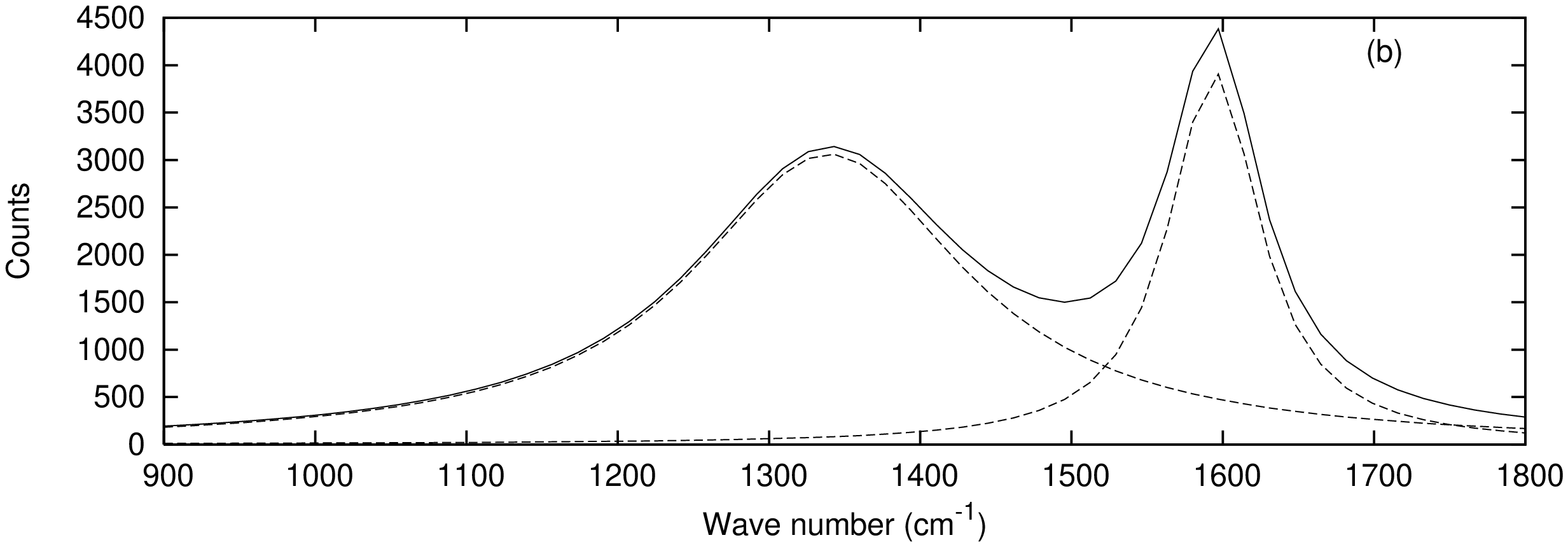}
\includegraphics[width=1.7in,angle=-90]{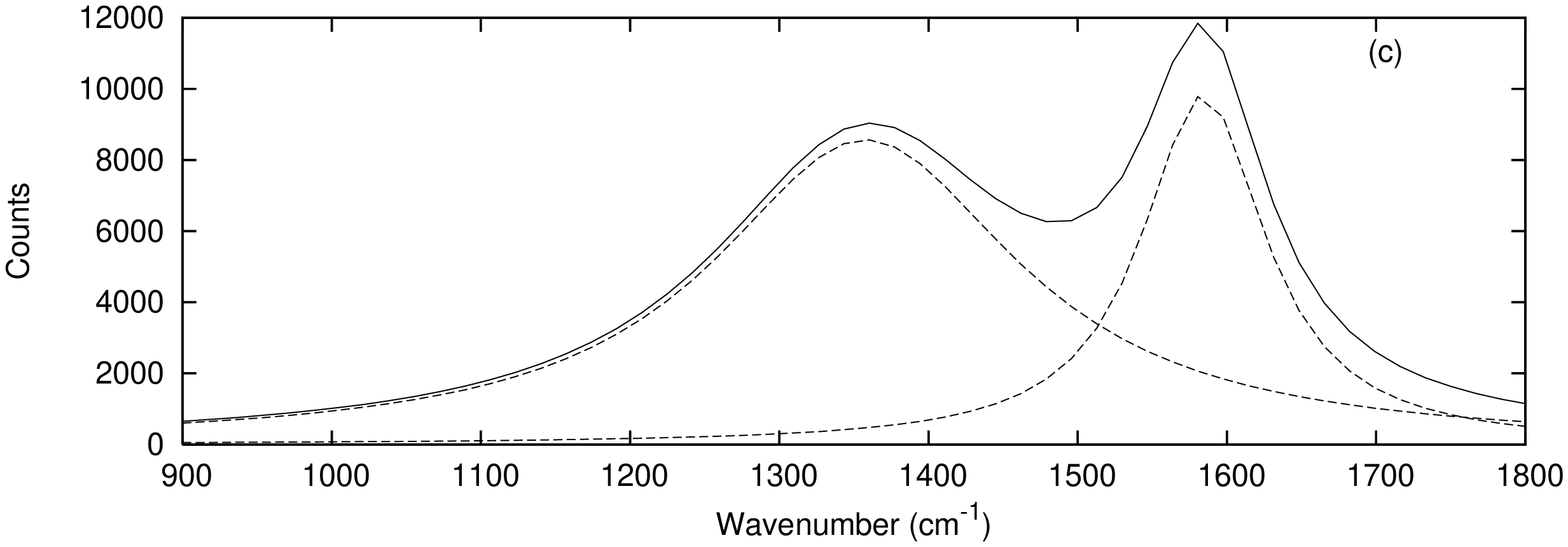}
\caption{Raman Spectra of DLC films deposited on quartz substrates
maintained at temperature (a) ${\rm 100^oC}$, (b) ${\rm 150^oC}$ and (c) 
${\rm 300^oC}$. The 
deconvulated curves represents the D and G peaks.}
\end{center}
\end{figure}
%%%%%%%%%%%%%%%%%%%%%%%%%%%%%%%%%%%%%%%%%%%%%%%%%%%%%%%%%%%%%%%%%%%%%%%%%%

%%%%%%%%%%%%%%%%%%%%%%%%%%%%%%%%%%%%%%%%%%%%%%%%%%%%%%%%%%%%%%%%%%%%%%%%%%%
\begin{figure}[b!!]
\begin{center}
\includegraphics[width=3in,angle=-90]{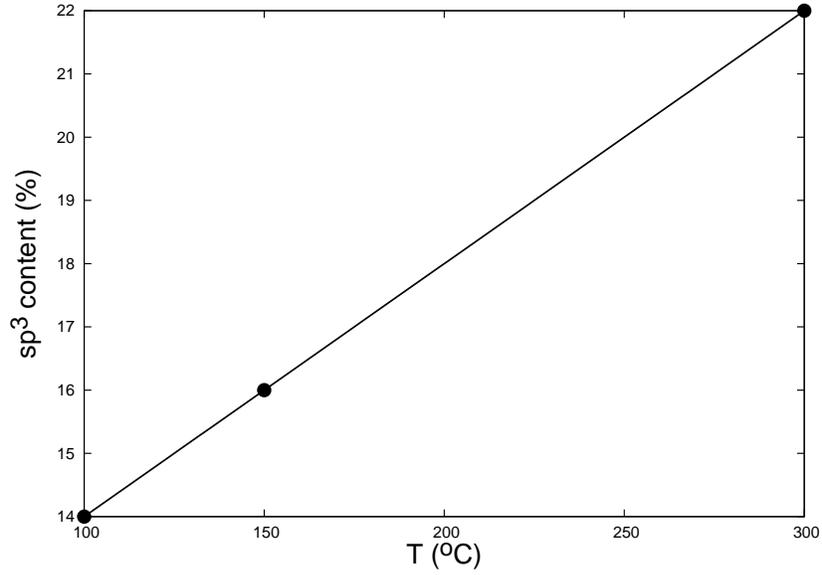}
\caption{The content of carbon atoms with ${\rm sp^3}$ bondings increases
linearly with temperature at which the quartz substrates were maintained.
}
\end{center}
\end{figure}
%%%%%%%%%%%%%%%%%%%%%%%%%%%%%%%%%%%%%%%%%%%%%%%%%%%%%%%%%%%%%%%%%%%%%%%%%%

%%%%%%%%%%%%%%%%%%%%%%%%%%%%%%%%%%%%%%%%%%%%%%%%%%%%%%%%%%%%%%%%%%%%%%%%%%%
\begin{figure}[t!!]
\begin{center}
\includegraphics[width=1.7in,angle=-90]{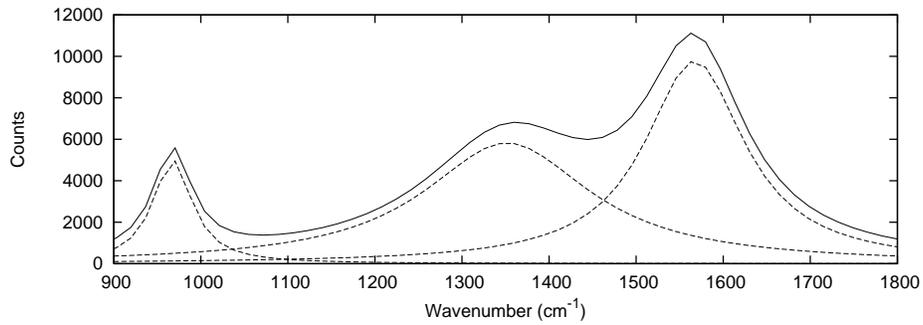}
\caption{Raman Spectra of DLC films deposited on Silicon substrates
maintained at ${\rm 300^oC}$. The 
deconvulated curves represents the D and G peaks. The peak between
900-1000${\rm cm^{-1}}$ is of silicon from the substrate.}
\end{center}
\end{figure}
%%%%%%%%%%%%%%%%%%%%%%%%%%%%%%%%%%%%%%%%%%%%%%%%%%%%%%%%%%%%%%%%%%%%%%%%%%
    
%%%%%%%%%%%%%%%%%%%%%%%%%%%%%%%%%%%%%%%%%%%%%%%%%%%%%%%%%%%%%%%%%%%%%%%%%%%
\begin{figure}[t!!]
\begin{center}
\includegraphics[width=3in,angle=-90]{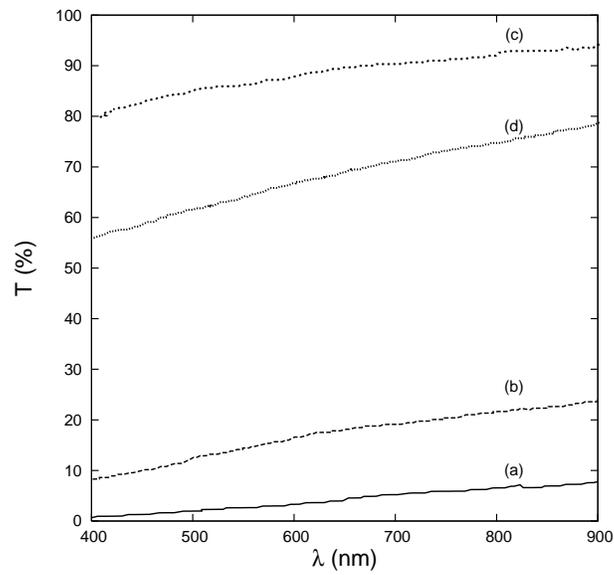}
\caption{The UV-visisble transmission spectra of DLC films deposited on 
quartz substrates maintained at temperature (a) room temperature, (b) 
${\rm 100^oC}$, (c) ${\rm 150^oC}$ and (d) ${\rm 300^oC}$.}
\end{center}
\end{figure}
%%%%%%%%%%%%%%%%%%%%%%%%%%%%%%%%%%%%%%%%%%%%%%%%%%%%%%%%%%%%%%%%%%%%%%%%%%

\end{document}